\documentclass[%
 reprint,
showpacs,
 amsmath,amssymb,
 aps, prl,
floatfix
]{revtex4-1}

\usepackage{graphicx}
\usepackage{epstopdf}
\usepackage{dcolumn}
\usepackage{bm}
\usepackage{makecell}
\usepackage{xcolor}

\begin{document}

\title{Stochastic Model of Breakdown Nucleation under Intense Electric Fields}
\author{Eliyahu Zvi Engelberg}
\author{Yinon Ashkenazy}
\author{Michael Assaf}
\affiliation{
  Racah Institute of Physics and
  the Center for Nanoscience and Nanotechnology, 
  Hebrew University of Jerusalem,
  Jerusalem 9190401, Israel
}
\date{\today}

\begin{abstract}
  Plastic response due to dislocation activity under intense electric fields
  is proposed as a source of breakdown.
  A model is formulated based on stochastic multiplication and arrest
  under the stress generated by the field.
  A critical transition in the dislocation population is suggested
  as the cause of protrusion formation leading to subsequent arcing.
  The model is studied using Monte Carlo simulations and theoretical analysis,
  yielding a simplified dependence of the breakdown rates
  on the electric field.
  These agree with experimental observations
  of field and temperature breakdown dependencies.
\end{abstract}

\maketitle

Various modern applications rely on maintaining high electric fields in a vacuum
between metallic electrodes \cite{clic_stage, boxman96, slade08, gai14}.
In such systems, arcing of current through the vacuum,
which leads to a field breakdown (BD), is a major failure mechanism.
Even when plasma formation is required \cite{keidar},
arcing nucleation and the mechanism leading up to it
play a critical role in system design.
Therefore,
arcing nucleation and vacuum BD are subjects of interest
in application and theory.

In most cases, plasma, created by particles emitted from the cathode,
leads to arcing \cite{boxman96, anders08, anders14}.
While the properties of arcing in relation to the interelectrode environment,
as well as the development of the cathode and anode surface states,
have been previously studied
and remain an active and important area of research,
the processes in the cathode leading up to BD nucleation
are not yet understood \cite{boxman96, gai14}.
It has been postulated that plastic damage to the cathode surface
plays a critical role in the nucleation process. 
Based on this assumption,
it was suggested that a BD can be related
to the mobility of defects within the solid
\footnote
{
  Contaminants and oxide layers can also lead to
  extrinsic BD nucleation.
  However, this paper discusses the rate of intrinsic BDs,
  which are observed after conditioning processes
  have removed sources for extrinsic BDs \cite{wuensch13}.
},
and thus the mean time to BD $\tau$
would show an exponential dependence on the stress $\sigma$.
Assuming that the leading contribution to $\sigma$
is due to Maxwell stress $\sigma = \epsilon_0 E^2 / 2$,
where $E$ is the applied field,
led to a $\ln\tau \sim E^2$ dependence.
This showed a good fit to a compilation of experimental data
on the BD rate (BDR $\sim 1/\tau$) versus
$E$ in metals at room temperature \cite{nordlund12}.

Understanding the observed limit is of general interest
and is important for the design of high gradient applications,
specifically in the proposed new CLIC project
in CERN \cite{grudiev09, gai14, clic_stage}.
This led to a concentrated effort to identify the mechanism
by which BDs are driven \cite{zadin14, vigonski15}.
The basic assumption was that an applied field
may cause a yield at the surface,
which would lead to the formation of a localized protrusion.
This protrusion would then enhance the electric current on the surface,
leading to heating and thus to plasma formation and arc nucleation.
However,
molecular dynamics and finite element simulations showed these processes
occurring only at $E \gtrsim$ 1 GV/m \cite{zadin14, vigonski15},
significantly more than the observed BD fields,
in the range of 100-200 MV/m \cite{grudiev09, gai14}.

It is well established that plasticity in metals close to the yield
is controlled by stochastic dislocation reactions \cite{Meyers_2002}. 
Individual crystals, too, deform via a sequence of discrete slip events,
as was demonstrated in the compression of micropillars \cite{Dimiduk_2009}.
The probability distribution of these events was measured \cite{dimiduk06}
and reproduced by discrete dislocation dynamics simulations
\cite{csikor07},
as well as mean-field models \cite{Dahmen_2012}.
Such systems demonstrate universal critical behavior characteristics of
a self-organized critical state controlled by a minimally stable cluster,
where in this case the cluster is a pinned dislocation arrangement
\cite{sethna_2011, Wiesenfeld_1987}.
A model reproducing this type of critical behavior
utilizes terms describing the kinetics of the mobile dislocation density $\rho$,
with nucleation at stress concentration sites on free surfaces, $\dot{\rho}_+$,
as well as their depletion, $\dot{\rho}_-$ \cite{Nix_Lee_2011}.
While not fully descriptive of the complex dislocation system,
this model successfully describes the nature and size dependence
of the observed stress-strain curves \cite{ryu13, nix_gao_2015}.

In a similar fashion, it has been shown, using a stochastic model,
that the correlated motion of dislocations
can lead to micron-sized surface protrusions,
when persistent slip bands, caused by cyclic stress,
break through the surface \cite{man09, levitin09}.
This was observed using SEM in fatigued samples exposed to high-cyclic stresses
\cite{goto08, laurent11}.
Samples exposed to strong electric fields, however,
do not show such prominent features \cite{gai14}.

Here we explore the possibility that the mechanism leading to arc nucleation
is a critical transition in the mobile dislocation population density
close to the surface.
We propose that, in a cathode subjected to an external electric field,
the dislocation density typically fluctuates around a stable level,
which depends on $E$.
However, at any point in time,
there is a finite probability that the density will reach a critical point,
beyond which it will increase deterministically.
Arc nucleation will then follow from the surface response
to the sudden localized increase in the surface dislocation density,
through a mechanism which we do not attempt to address at this stage. 

To explore this option,
we employ a zero-dimensional mean-field model
to describe the kinetics of creation and depletion of mobile dislocations
in a single slip plane,
neglecting interactions between slip planes
and the spatial variation of the mobile dislocation density within one plane.
In this model, mobile dislocations nucleate at existing sources,
and their depletion is due to collisions with obstacles.
We formulate the problem in terms of a birth-death master equation
\cite{gardiner04} for the mobile dislocation population.
This formalism is used to calculate an explicit analytical expression
for the BDR in one slip plane as a function of $E$,
which agrees well with kinetic Monte Carlo (KMC) simulations.
The resulting model is unique in that, for the first time,
it treats a BD in metals as a critical transition,
due to the stochastic evolution of dislocations under an external field.
In contrast with linear evolution models \cite{nordlund12, zadin14, vigonski15},
our model predicts an eventual BD
without requiring observable pre-BD surface features,
whose absence in the microscopy of post-BD samples
has posed a long-standing problem \cite{gai14}.

We calibrate the model for oxygen-free high-conductivity Cu,
due to the availability of experimental data,
as it is used in the CERN CLIC Collaboration,
developing the next-generation linear collider accelerators.
The physical parameters that are unknown are found
by fitting the results generated by the model to experimental observations,
including the temperature and $E$ dependence of the BDR.
Following this calibration,
the model yields a quantitative agreement with the observed experimental BDRs,
without making additional assumptions about the physical characteristics
of the system,
such as postulating the existence of specific surface or subsurface features.

\emph{Deterministic description.}\textemdash
A simplified kinetic model is based on the average creation and depletion rates
$\dot{\rho}_+$ and $\dot{\rho}_-$, respectively \cite{Nix_Lee_2011}.
For these, we assume that the kinetics are described within slip planes
limited by dislocation cells on the order of 10 $\mu$m.
Dislocations nucleate at sources,
whose density depends on the number of mobile dislocations,
at a rate depending on $E$.
They are depleted by interactions with other mobile dislocations
and existing defects.
Thus, the deterministic dynamics of the mobile dislocation density are given by
\begin{align}
  \label{eq_rho}
  & \dot{\rho} = \dot{\rho}_+ - \dot{\rho}_- \\
  & \dot{\rho}_+ = B_1(\rho + c) \sigma^2 e^{\alpha\sigma} \, ; \quad
  \dot{\rho}_- = b_2\sigma\rho(\rho + c), \nonumber 
\end{align}
with $\sigma = A_1 + a_2\rho$.
The constants $A_1$, $a_2$, $B_1$, $b_2$, $c$,
and $\alpha$ depend on the system parameters
and are independent of $E$,
except for $A_1 \sim E^2$.
The derivation of these functional forms
and the relation of the effective constants to the
appropriate physical parameters
is described in Supplemental Material;
see also Fig. S1 \cite{supplemental_material}.
Henceforth,
all analytical and numerical results are presented for fitted
parameters (see below).

Below a critical field $E_c$ (see below),
the equation $\dot{\rho} = 0$ yields two solutions:
$\rho_*$ and $\rho_c$, where $\rho_* < \rho_c$.
For $\rho_* < \rho < \rho_c$, we have $\dot{\rho}_- > \dot{\rho}_+$,
and the dislocation density deterministically decreases back to $\rho_*$.
Therefore, $\rho_*$ is an attracting fixed point of Eq. (\ref{eq_rho}),
while $\rho_c$ is a repelling point.
That is, if $\rho$ is larger than $\rho_c$, it will increase indefinitely,
leading to a subsequent BD.

Sufficiently below $E_c$, where $\rho_* \ll A_1 / a_2 \ll \rho_c$,
we can find analytical expressions for $\rho_*$ and $\rho_c$,
which yield $\rho_* = (B_1 A_1 / b_2) e^{\alpha A_1}$ and 
$\rho_c = (\alpha a_2)^{-1} \ln[b_2 / (B_1 a_2)]$.

\begin{figure}
  \centering
  \includegraphics[width = 8.5cm]{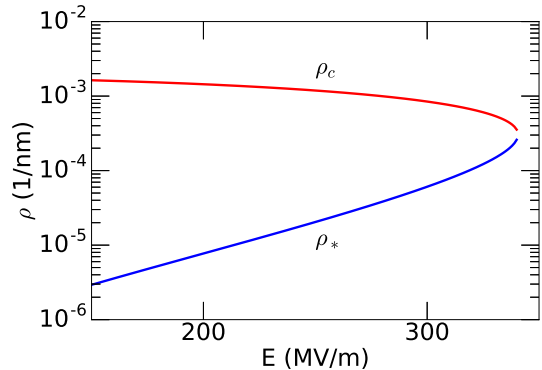} \\[-5ex]
  \caption
  {
    \label{fig_rho_edges}
    Fixed points of the dynamical equations for $\rho$,
    attracting ($\rho_*$) and repelling ($\rho_c$),
    as functions of $E$,
    demonstrating the existence of a bifurcation point.
  }
\end{figure}

The values of $\dot{\rho}_+$ and $\dot{\rho}_-$
approach each other as $E$ increases,
as shown in Fig. \ref{fig_rho_edges}.
At $E = E_c$, $\rho_* = \rho_c$,
and thus for $E \geq E_c$
the system progresses deterministically to a BD.
Notably, for $E < E_c$,
reaching $\rho_c$ is a fluctuation-driven stochastic event,
leading to an $E$-dependent BDR.

\emph{Stochastic model.}\textemdash
To incorporate fluctuations,
we model the evolution of the mobile dislocation density $\rho$
as a birth-death Markov process \cite{gardiner04}.
The rates $\dot{\rho}_+$ and $\dot{\rho}_-$ represent the probability per unit
time that $\rho$ will increase or decrease,
respectively, by $\Delta\rho = 0.1 \, \mu\text{m}^{-1}$,
corresponding to one dislocation per cell.
For one single slip plane, we define $n = \rho / \Delta\rho$
as the instantaneous number of mobile dislocations per cell.
By defining $A_2 = a_2 n_c \Delta\rho$,
$B_2 = b_2 n_c \Delta\rho$,
and $C = c / (n_c \Delta\rho)$,
where $n_c = \rho_c / \Delta\rho $,
we find that the microscopic birth and death rates as a function of $n$ are,
respectively,
\begin{equation}
  \lambda_n = B_1(n + n_c C) \sigma^2 e^{\alpha\sigma} \label{eq_rates} ; \;
  \mu_n = (B_2 n / n_c) (n + n_c C) \sigma,
\end{equation}
with $\sigma(n) = A_1 + A_2 n / n_c$
\footnote
{
  The parameters have been expressed in this form so that we can later
  conveniently apply a WKB-like perturbation theory with respect to $n_c \gg 1$
  \cite{dykman94}.
}.
The stochastic dynamics are governed by the master equation
\begin{equation}
  \frac{\partial P_n(t)}{\partial t} = \lambda_{n - 1} P_{n - 1}(t)
  + \mu_{n + 1} P_{n + 1}(t) - (\lambda_n + \mu_n) P_n(t),
  \label{eq_master_probability}
\end{equation}
describing the evolution of the probability $P_n(t)$ of finding $n$
mobile dislocations per cell at time $t$ \cite{gardiner04}.

In order to find the BDR, we write a recursive equation for $T_n$,
the mean time it takes to reach BD
starting from $n$ mobile dislocations \cite{gardiner04, assaf17}:
\begin{equation}
  T_n = \frac{\lambda_n}{\lambda_n + \mu_n} T_{n + 1}
  + \frac{\mu_n}{\lambda_n + \mu_n} T_{n - 1} + \frac{1}{\lambda_n + \mu_n}
  \label{eq_mean_passage_time}
\end{equation}
where $(\lambda_n + \mu_n)^{-1}$ is the average time it takes to jump
from $n$ to $n\pm 1$.
We solve this equation with an absorbing boundary at $n_c$
and a reflecting wall at $n = 0$,
such that $T_{n_c} = 0$ and $T_n'(n = 0) = 0$.
Starting from the vicinity of $n = 0$, $\tau$ is given by
\footnote
{
  Note that, since $n_* = O(1)$ is an attracting fixed point
  of Eq. (\ref{eq_rho}),
  starting from any $n = O(1)$ in Eq. (\ref{eq_tau_exact})
  leads to the same quantitative result.
}
\begin{equation}
  \tau = T_n = \sum_{i = n}^{n_c} \phi_i
  \left( \sum_{j = 0}^i \frac{1}{\lambda_j \phi_j} \right),
  \label{eq_tau_exact}
\end{equation}
with $\phi_n = \prod_{m = 1}^{n} \mu_m / \lambda_m$ \cite{gardiner04}.

\begin{figure}
  \includegraphics[width = 8.5cm]{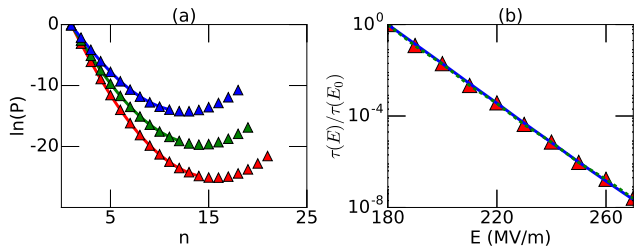} \\[-3ex]
  \caption
  {
    \label{fig_qsd_mbt}
    Analytical versus simulated values:
    (a) Probability of being in state $n$, at $t \ll \tau$,
    as calculated from the
    analytical expression [Eq. (\ref{eq_qsd})] and numerical simulation.
    The lines, from bottom to top, are for $E$ = 200, 230 and 260 MV/m.
    The simulation points include measurements of probability for $n > n_c$,
    above the QSD validity regime.
    (b) $\tau$ normalized by $\tau(E_0 = 180\,\text{MV/m})$
    as a function of $E$,
    calculated using the metastable approximation
    [Eq. (\ref{eq_tau_recursive_final}), solid line],
    the exact formula [Eq. (\ref{eq_tau_exact}), dashed line],
    and the simulation (triangles).
  }
\end{figure}

In the rest of the Letter,
we focus on the regime where the critical number of mobile dislocations,
needed for a BD, satisfies $n_c \gg 1$.
In this regime, the expression for $\tau$ can be simplified.
Assuming \emph{a priori} that $\tau$ is exponentially large in $n_c$,
we can employ the metastability ansatz $P_n(t) \simeq \pi_n e^{-t / \tau}$
\footnote
{
  Here, once the system has already settled into a long-lived
  metastable state centered about $n_*$,
  the metastable distribution slowly decays at a rate of $1 / \tau$,    
  representing the BD probability.
},
where $\pi_n$ is the \emph{quasistationary distribution} (QSD)
\cite{dykman94, assaf06, escudero09, assaf10}.
Substituting this into Eq. (\ref{eq_master_probability})
and neglecting the exponentially small term proportional to $1 / \tau$,
we have 
$\lambda_{n-1} \pi_{n-1} + \mu_{n+1} \pi_{n+1} - (\lambda_n + \mu_n) \pi_n = 0$, 
whose solution, for $n \leq n_c$
\footnote
{
  Here we assume $\pi_{n > n_c} \simeq 0$ at $t \ll \tau$.
},
satisfies \cite{gardiner04}
\begin{equation}
  \pi_n = \pi_0 \prod_{m = 1}^{n} \frac{\lambda_{m - 1}}{\mu_m}.
  \label{eq_qsd}
\end{equation}
This solution is shown in Fig. \ref{fig_qsd_mbt}(a),
where $\pi_0$ is found via normalization \cite{gardiner04}.
Using Eq. (\ref{eq_master_probability}) for $n = n_c + 1$
and the metastability ansatz,
we thus have $\tau \simeq (\lambda_{n_c}\pi_{n_c})^{-1}$
\cite{dykman94, assaf06, escudero09, assaf10}.
Using Eqs. (\ref{eq_rates}) and (\ref{eq_qsd}),
expanding in $n_c \gg 1$,
and applying the Stirling approximation up to subleading order,
$\tau$ gives way to a WKB-like solution \cite{dykman94}
\begin{equation}
  \tau = \mathcal{A} e^{n_c \Delta S}, \label{eq_tau_recursive_final}
\end{equation}
where
\begin{align}
  \label{eq_terms_in_exponent}
  \Delta S &= \ln\frac{B_2}{A_1 B_1} -
  \alpha A_1 \left(1 + \frac{1}{2\eta}\right) - (\eta + 1)
  \ln\left(1 + \frac{1}{\eta}\right), \\
  \mathcal{A} &= \sqrt{\frac{2\pi}{n_c}}
  \frac{e^{-\alpha A_1 [1 + (1/2\eta)]}}{A_1^2 B_1 C}
  \left(1 + \frac{1}{\eta}\right)^{-1/2}, \nonumber
\end{align}
and $\eta = A_1 / A_2$.
Here $n_c \Delta S$ can be viewed as a barrier to a BD
\footnote
{
  When $E \simeq E_c$,
  the term $A_1 \sim E^2$ dominates the exponent.
  However, in the experimental regime of interest,
  where the BD is fluctuation driven,
  all terms in Eq. (\ref{eq_terms_in_exponent})
  are of similar magnitude.
}.
Note that, in the experimentally relevant electric-field range,
we observe that
\begin{equation}
\tau \sim \exp[\gamma\left(1 - E / E_0)\right] \label{eq_linear_ln_tau}
\end{equation}
with $E_0$ a reference field
and $\gamma$ a dimensionless constant independent of $E$,
as demonstrated in Fig. \ref{fig_qsd_mbt}(b)
\footnote
{
  While the $E^2$ provides a reasonable fit
  to experimentally available data at room temperature \cite{nordlund12},
  a linear dependence provides a better approximation
  over a wide range of electric fields and temperatures
  \cite{supplemental_material}.
}.

The analytical results were compared to a KMC simulation,
tracking the time evolution of $n$.
In this simulation,
$n\to n\pm 1$ changes randomly
using the transition rates in Eq. (\ref{eq_rates}),
with the time elapsed between changes determined
using an exponential distribution with mean $(\lambda_n + \mu_n)^{-1}$.
The numerically estimated QSD and $\tau$ agree with the analytical solution;
see Figs. \ref{fig_qsd_mbt} and S2 \cite{supplemental_material}.

\emph{Parameter range.}\textemdash
The model includes six constants
which depend on the material properties and on specific mechanisms that 
control the reactions of dislocations to the applied field
\cite{supplemental_material}.
These constants depend on four unknown parameters:
(i) $\beta$,
relating the stress at the nucleation sites to $E$;
(ii) $\kappa$, representing temperature-independent factors affecting nucleation,
such as the nucleation attempt frequency and activation entropy;
(iii) the activation energy $E_a$; and
(iv) the activation volume $\Omega$ for releasing new mobile dislocations.

\begin{figure}
  \includegraphics[width = 8.5cm]{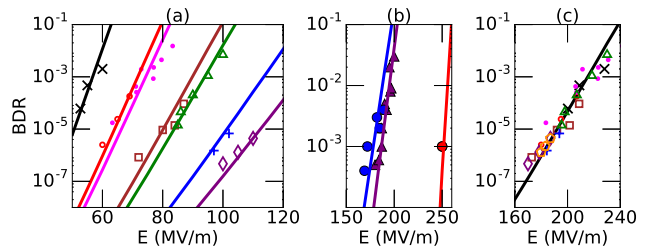} \\[-3ex]
  \caption
  {
    \label{fig_fits}
    Experimental BDRs with fitted theoretical lines
    using Eq. (\ref{eq_tau_recursive_final}):
    (a) BDR versus $E$ for various Cu accelerating structures
    \cite{grudiev09}.
    (b) BDR variation with $E$ at room temperature (two lines on the left)
    and at 45 K (line on the right) \cite{cahill17}.
    (c) BDR versus $E$ for various Cu accelerating structures
    \cite{grudiev09, wuensch17},
    with $E$ rescaled so that all measurements are fitted with
    $\beta$ = 4.8.
  }
\end{figure}

The values of these parameters can be found and validated by a comparison to
experimental measurements of the BDR
as a function of $E$ for different structures \cite{grudiev09},
shown in Fig. \ref{fig_fits}(a),
and for one structure at both room temperature and 45 K \cite{cahill17},
shown in Fig. \ref{fig_fits}(b).
The results include measurements from various geometries,
leading to a significant variation in the local field
at the BD site \cite{grudiev09}. 
As this translates to a variation in $\beta$ only,
results were scaled using $\beta$
to those of a reference set \cite{wuensch17},
so that all sets produce identical BDRs at $E$ = 180 MV/m.
This led to a single normalized 
data set shown in Fig. \ref{fig_fits}(c).
As normalization was done for each structure at 300 K,
results at 45 K are presented for the rescaled $E$ = 250 MV/m,
rather than the measured 300 MV/m.

\begin{figure}
  \centering
  \includegraphics[width = 8.5cm]{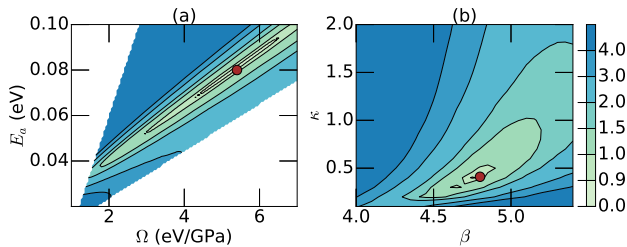} \\[-3ex]
  \caption
  {
    \label{fig_colormaps}
    Fit of free model parameters: 
    (a) LSQ parameter as a function of $\Omega$ and $E_a$,
    with $\beta$ = 4.8 and $\kappa$ = 0.41.
    (b) LSQ parameter as a function of $\beta$ and $\kappa$,
    with $\Omega$ = 5.4 eV/GPa and $E_a$ = 0.08 eV.
  }
\end{figure}

The parameter evaluation was done in two steps:
First,
a fitting was performed for the reference set \cite{wuensch17}
and the data from the temperature-varied structure \cite{cahill17},
using a least square (LSQ) fit demanding
(i) consistency with the experimental values of 
$\tau(T = 300\,\text{K}, E = 180\,\text{Mv/m})$,
$\tau(T = 45\,\text{K}, E = 300\,\text{Mv/m})$,
and $\gamma(T = 300\,\text{K})$ of Eq. (\ref{eq_linear_ln_tau}),
(ii) consistency with theoretical estimates of $E_a \gtrsim 0.1$, and
(iii) validity of the approximation $n_c \gg 1$
at the corresponding temperatures and fields as in (i)
\footnote
{
  $E_a \simeq 0.1$ is the lowest estimate to date of the activation energy
  of mobile dislocation nucleation \cite{zhu08}.
}.
Next, the rest of the data sets \cite{grudiev09}
were used to validate the quality of the fit.
Two cross sections of the resulting LSQ fit in the parameter space
($\Omega$, $E_a$, $\beta$, $\kappa$)
are plotted in Fig. \ref{fig_colormaps}.
We find that the LSQ parameter has a minimum at
$\beta$ = 4.8$\pm$0.1, $\kappa$ = 0.41$\pm$0.02,
$\Omega$ = 5.4$\pm$0.2 eV/GPa, and $E_a$ = 0.08$\pm$0.002 eV,
marked on the graphs in Fig. \ref{fig_colormaps}.

Our results for $E_a$ are consistent with mobile dislocation nucleation
from preexisting sources \cite{zhu08},
significantly lower than the activation energy for nucleation in pristine
crystalline structures \cite{bonneville88, couteau11, ryu11}.
Furthermore, the activation volume $\Omega$ = 55$b^3$, with $b$ the Burgers vector,
is within the experimental range $10b^3 < \Omega < 124b^3$ \cite{zhu08, couteau11}.

\emph{Discussion.}\textemdash
The model can be consistently fitted to
all available experimental data sets,
with a single free parameter $\beta$ adjusted to account
for the geometrical difference between experimental structures.
Thus,
the model allows us to make predictions for BDRs over
a wide range of physical parameters
beyond those of the current measurements,
as demonstrated in Fig. \ref{fig_fits}.

According to the proposed model,
BD nucleation is preceded by a critical increase
in the number of mobile dislocations.
This can create an early-warning signal for imminent BD
through the monitoring of characteristic fluctuations \cite{scheffer09},
which includes indirect measurements
such as thermionic current emissions,
or direct measurements of acoustic signals
from increased fluctuations in the mobile dislocation populations
\cite{weiss07}.
Furthermore, it is expected that the standard deviation of the QSD,
representing the typical fluctuations of the pre-BD
mobile dislocation population,
will increase significantly as the BD is approached
\cite{supplemental_material}.

Conversely, our model does not depend upon the appearance
of an observable surface protrusion before a BD.
This is in line with the fact that no observable sub-BD features
have been observed in metallic electrodes
exposed to a strong electric field \cite{gai14}.
BD sites are characterized by a large crater created by the arc
\cite{boxman96},
obliterating any remains of possible pre-BD features.
Such features, however,
should have been found further away from the BD site if they existed.

An understanding of the mechanism which leads to BD nucleation
can facilitate the development of better design of electrodes,
focusing on limiting the nucleation and mobility of dislocations,
in order to lower the BDR.
It is well established
that, in order to stabilize a significant field,
an electrode has to undergo \emph{conditioning}
via a series of field exposures
and BDs at lower fields \cite{degiovanni16}.
Conditioning includes both an initial extrinsic process 
(resulting from the removal of contaminates)
and a long-term intrinsic process 
resulting from modifications of the electrode.
In line with our model, this comes about as a result
of surface hardening \cite{wuensch13}.
In addition, by controlling the dislocation mobility,
our model offers a direction for improving the electrode performance.

In conclusion,
we present a model describing BD nucleation
as a stochastic process driven by the creation and depletion of dislocations
within the electrode.
BD nucleation in this case is a result of a critical transition
in the mobile dislocation population density.
The model was formulated using a set of parameters
describing known material properties
and unknown parameters describing interactions
specific to the response of the dislocation population to the applied field.
Measurements in various fields and temperatures
were used to fit the parameters and validate the model.
This model is unique, as it does not require pre-BD features,
and offers a simple intrinsic mechanism
for a BD at fields lower than the deterministic limit.
Establishing such a model may provide opportunities
for improving the design of future electrodes,
aiming to limit the dislocation mobility,
as well as offer ways to identify pre-BD early-warning signals
through the evolution of the dislocation population.

\begin{acknowledgments}
  We acknowledge K. Nordlund, F. Djurabekova, and W. Wuensch
  for  helpful discussions and providing data for Fig. \ref{fig_fits}.
  We acknowledge funding from the PAZI foundation.
\end{acknowledgments}

\bibliography{main}

\begin{thebibliography}{60}%
\makeatletter
\providecommand \@ifxundefined [1]{%
 \@ifx{#1\undefined}
}%
\providecommand \@ifnum [1]{%
 \ifnum #1\expandafter \@firstoftwo
 \else \expandafter \@secondoftwo
 \fi
}%
\providecommand \@ifx [1]{%
 \ifx #1\expandafter \@firstoftwo
 \else \expandafter \@secondoftwo
 \fi
}%
\providecommand \natexlab [1]{#1}%
\providecommand \enquote  [1]{``#1''}%
\providecommand \bibnamefont  [1]{#1}%
\providecommand \bibfnamefont [1]{#1}%
\providecommand \citenamefont [1]{#1}%
\providecommand \href@noop [0]{\@secondoftwo}%
\providecommand \href [0]{\begingroup \@sanitize@url \@href}%
\providecommand \@href[1]{\@@startlink{#1}\@@href}%
\providecommand \@@href[1]{\endgroup#1\@@endlink}%
\providecommand \@sanitize@url [0]{\catcode `\\12\catcode `\$12\catcode
  `\&12\catcode `\#12\catcode `\^12\catcode `\_12\catcode `\%12\relax}%
\providecommand \@@startlink[1]{}%
\providecommand \@@endlink[0]{}%
\providecommand \url  [0]{\begingroup\@sanitize@url \@url }%
\providecommand \@url [1]{\endgroup\@href {#1}{\urlprefix }}%
\providecommand \urlprefix  [0]{URL }%
\providecommand \Eprint [0]{\href }%
\providecommand \doibase [0]{http://dx.doi.org/}%
\providecommand \selectlanguage [0]{\@gobble}%
\providecommand \bibinfo  [0]{\@secondoftwo}%
\providecommand \bibfield  [0]{\@secondoftwo}%
\providecommand \translation [1]{[#1]}%
\providecommand \BibitemOpen [0]{}%
\providecommand \bibitemStop [0]{}%
\providecommand \bibitemNoStop [0]{.\EOS\space}%
\providecommand \EOS [0]{\spacefactor3000\relax}%
\providecommand \BibitemShut  [1]{\csname bibitem#1\endcsname}%
\let\auto@bib@innerbib\@empty
\bibitem [{\citenamefont {Burrows}\ \emph {et~al.}(2016)\citenamefont
  {Burrows}, \citenamefont {Lebrun}, \citenamefont {Linssen}, \citenamefont
  {Schulte}, \citenamefont {Sicking}, \citenamefont {Stapnes},\ and\
  \citenamefont {Thomson}}]{clic_stage}%
  \BibitemOpen
  \bibfield  {author} {\bibinfo {author} {\bibfnamefont {P.}~\bibnamefont
  {Burrows}}, \bibinfo {author} {\bibfnamefont {P.}~\bibnamefont {Lebrun}},
  \bibinfo {author} {\bibfnamefont {L.}~\bibnamefont {Linssen}}, \bibinfo
  {author} {\bibfnamefont {D.}~\bibnamefont {Schulte}}, \bibinfo {author}
  {\bibfnamefont {E.}~\bibnamefont {Sicking}}, \bibinfo {author} {\bibfnamefont
  {S.}~\bibnamefont {Stapnes}}, \ and\ \bibinfo {author} {\bibfnamefont
  {M.}~\bibnamefont {Thomson}},\ }\href@noop {} {\emph {\bibinfo {title}
  {Updated baseline for a staged Compact Linear Collider}}},\ \bibinfo {type}
  {Tech. Rep.}\ \bibinfo {number} {CERN-2016-004}\ (\bibinfo  {institution}
  {CERN},\ \bibinfo {address} {Geneva},\ \bibinfo {year} {2016})\BibitemShut
  {NoStop}%
\bibitem [{\citenamefont {Boxman}\ \emph {et~al.}(1996)\citenamefont {Boxman},
  \citenamefont {Sanders},\ and\ \citenamefont {Martin}}]{boxman96}%
  \BibitemOpen
  \bibinfo {editor} {\bibfnamefont {R.~L.}\ \bibnamefont {Boxman}}, \bibinfo
  {editor} {\bibfnamefont {D.~M.}\ \bibnamefont {Sanders}}, \ and\ \bibinfo
  {editor} {\bibfnamefont {P.~J.}\ \bibnamefont {Martin}},\ eds.,\ \href@noop
  {} {\emph {\bibinfo {title} {Handbook of Vacuum Arc Science and Technology:
  Fundamentals and Applications}}}\ (\bibinfo  {publisher} {Noyes},\ \bibinfo
  {address} {Park Ridge, NJ},\ \bibinfo {year} {1996})\BibitemShut {NoStop}%
\bibitem [{\citenamefont {Slade}(2008)}]{slade08}%
  \BibitemOpen
  \bibfield  {author} {\bibinfo {author} {\bibfnamefont {P.~G.}\ \bibnamefont
  {Slade}},\ }\href@noop {} {\emph {\bibinfo {title} {The Vacuum Interrupter:
  Theory, Design, and Application}}}\ (\bibinfo  {publisher} {CRC},\ \bibinfo
  {address} {Boca Raton, FL},\ \bibinfo {year} {2008})\BibitemShut {NoStop}%
\bibitem [{\citenamefont {Gai}(2014)}]{gai14}%
  \BibitemOpen
  \bibinfo {editor} {\bibfnamefont {W.}~\bibnamefont {Gai}},\ ed.,\ \href@noop
  {} {\emph {\bibinfo {title} {High Gradient Accelerating Structure}}}\
  (\bibinfo  {publisher} {World Scientific},\ \bibinfo {address} {Singapore},\
  \bibinfo {year} {2014})\BibitemShut {NoStop}%
\bibitem [{\citenamefont {Teel}\ \emph {et~al.}(2017)\citenamefont {Teel},
  \citenamefont {Shashurin}, \citenamefont {Fang},\ and\ \citenamefont
  {Keidar}}]{keidar}%
  \BibitemOpen
  \bibfield  {author} {\bibinfo {author} {\bibfnamefont {G.}~\bibnamefont
  {Teel}}, \bibinfo {author} {\bibfnamefont {A.}~\bibnamefont {Shashurin}},
  \bibinfo {author} {\bibfnamefont {X.}~\bibnamefont {Fang}}, \ and\ \bibinfo
  {author} {\bibfnamefont {M.}~\bibnamefont {Keidar}},\ }\href {\doibase
  10.1063/1.4974004} {\bibfield  {journal} {\bibinfo  {journal} {J. Appl.
  Phys.}\ }\textbf {\bibinfo {volume} {121}},\ \bibinfo {pages} {023303}
  (\bibinfo {year} {2017})}\BibitemShut {NoStop}%
\bibitem [{\citenamefont {Anders}(2008)}]{anders08}%
  \BibitemOpen
  \bibfield  {author} {\bibinfo {author} {\bibfnamefont {A.}~\bibnamefont
  {Anders}},\ }\href@noop {} {\emph {\bibinfo {title} {Cathodic Arcs: From
  Fractal Spots to Energetic Condensation}}},\ \bibinfo {series} {Springer
  Series on Atomic, Optical, and Plasma Physics}, Vol.~\bibinfo {volume} {50}\
  (\bibinfo  {publisher} {Springer},\ \bibinfo {address} {New York},\ \bibinfo
  {year} {2008})\BibitemShut {NoStop}%
\bibitem [{\citenamefont {Anders}(2014)}]{anders14}%
  \BibitemOpen
  \bibfield  {author} {\bibinfo {author} {\bibfnamefont {A.}~\bibnamefont
  {Anders}},\ }\href@noop {} {\bibfield  {journal} {\bibinfo  {journal} {Surf.
  Coat. Technol.}\ }\textbf {\bibinfo {volume} {257}},\ \bibinfo {pages} {308}
  (\bibinfo {year} {2014})}\BibitemShut {NoStop}%
\bibitem [{Note1()}]{Note1}%
  \BibitemOpen
  \bibinfo {note} {Contaminants and oxide layers can also lead to extrinsic BD
  nucleation. However, this paper discusses the rate of intrinsic BDs, which
  are observed after conditioning processes have removed sources for extrinsic
  BDs \cite {wuensch13}.}\BibitemShut {Stop}%
\bibitem [{\citenamefont {Nordlund}\ and\ \citenamefont
  {Djurabekova}(2012)}]{nordlund12}%
  \BibitemOpen
  \bibfield  {author} {\bibinfo {author} {\bibfnamefont {K.}~\bibnamefont
  {Nordlund}}\ and\ \bibinfo {author} {\bibfnamefont {F.}~\bibnamefont
  {Djurabekova}},\ }\href@noop {} {\bibfield  {journal} {\bibinfo  {journal}
  {Phys. Rev. Accel. Beams}\ }\textbf {\bibinfo {volume} {15}},\ \bibinfo
  {pages} {071002} (\bibinfo {year} {2012})}\BibitemShut {NoStop}%
\bibitem [{\citenamefont {Grudiev}\ \emph {et~al.}(2009)\citenamefont
  {Grudiev}, \citenamefont {Calatroni},\ and\ \citenamefont
  {Wuensch}}]{grudiev09}%
  \BibitemOpen
  \bibfield  {author} {\bibinfo {author} {\bibfnamefont {A.}~\bibnamefont
  {Grudiev}}, \bibinfo {author} {\bibfnamefont {S.}~\bibnamefont {Calatroni}},
  \ and\ \bibinfo {author} {\bibfnamefont {W.}~\bibnamefont {Wuensch}},\
  }\href@noop {} {\bibfield  {journal} {\bibinfo  {journal} {Phys. Rev. Accel.
  Beams}\ }\textbf {\bibinfo {volume} {12}},\ \bibinfo {pages} {102001}
  (\bibinfo {year} {2009})}\BibitemShut {NoStop}%
\bibitem [{\citenamefont {Zadin}\ \emph {et~al.}(2014)\citenamefont {Zadin},
  \citenamefont {Pohjonen}, \citenamefont {Aabloo}, \citenamefont {Nordlund},\
  and\ \citenamefont {Djurabekova}}]{zadin14}%
  \BibitemOpen
  \bibfield  {author} {\bibinfo {author} {\bibfnamefont {V.}~\bibnamefont
  {Zadin}}, \bibinfo {author} {\bibfnamefont {A.}~\bibnamefont {Pohjonen}},
  \bibinfo {author} {\bibfnamefont {A.}~\bibnamefont {Aabloo}}, \bibinfo
  {author} {\bibfnamefont {K.}~\bibnamefont {Nordlund}}, \ and\ \bibinfo
  {author} {\bibfnamefont {F.}~\bibnamefont {Djurabekova}},\ }\href@noop {}
  {\bibfield  {journal} {\bibinfo  {journal} {Phys. Rev. Accel. Beams}\
  }\textbf {\bibinfo {volume} {17}},\ \bibinfo {pages} {103501} (\bibinfo
  {year} {2014})}\BibitemShut {NoStop}%
\bibitem [{\citenamefont {Vigonski}\ \emph {et~al.}(2015)\citenamefont
  {Vigonski}, \citenamefont {Veske}, \citenamefont {Aabloo}, \citenamefont
  {Djurabekova},\ and\ \citenamefont {Zadin}}]{vigonski15}%
  \BibitemOpen
  \bibfield  {author} {\bibinfo {author} {\bibfnamefont {S.}~\bibnamefont
  {Vigonski}}, \bibinfo {author} {\bibfnamefont {M.}~\bibnamefont {Veske}},
  \bibinfo {author} {\bibfnamefont {A.}~\bibnamefont {Aabloo}}, \bibinfo
  {author} {\bibfnamefont {F.}~\bibnamefont {Djurabekova}}, \ and\ \bibinfo
  {author} {\bibfnamefont {V.}~\bibnamefont {Zadin}},\ }\href@noop {}
  {\bibfield  {journal} {\bibinfo  {journal} {Appl. Math. and Comput.}\
  }\textbf {\bibinfo {volume} {267}},\ \bibinfo {pages} {476} (\bibinfo {year}
  {2015})}\BibitemShut {NoStop}%
\bibitem [{\citenamefont {Meyers}\ \emph {et~al.}(2002)\citenamefont {Meyers},
  \citenamefont {Benson}, \citenamefont {V\"ohringer}, \citenamefont {Kad},
  \citenamefont {Xue},\ and\ \citenamefont {Fu}}]{Meyers_2002}%
  \BibitemOpen
  \bibfield  {author} {\bibinfo {author} {\bibfnamefont {M.~A.}\ \bibnamefont
  {Meyers}}, \bibinfo {author} {\bibfnamefont {D.~J.}\ \bibnamefont {Benson}},
  \bibinfo {author} {\bibfnamefont {O.}~\bibnamefont {V\"ohringer}}, \bibinfo
  {author} {\bibfnamefont {B.~K.}\ \bibnamefont {Kad}}, \bibinfo {author}
  {\bibfnamefont {Q.}~\bibnamefont {Xue}}, \ and\ \bibinfo {author}
  {\bibfnamefont {H.-H.}\ \bibnamefont {Fu}},\ }\href@noop {} {\bibfield
  {journal} {\bibinfo  {journal} {Mater. Sci. Eng., A}\ }\textbf {\bibinfo
  {volume} {322}},\ \bibinfo {pages} {194} (\bibinfo {year}
  {2002})}\BibitemShut {NoStop}%
\bibitem [{\citenamefont {Uchic}\ \emph {et~al.}(2009)\citenamefont {Uchic},
  \citenamefont {Shade},\ and\ \citenamefont {Dimiduk}}]{Dimiduk_2009}%
  \BibitemOpen
  \bibfield  {author} {\bibinfo {author} {\bibfnamefont {M.~D.}\ \bibnamefont
  {Uchic}}, \bibinfo {author} {\bibfnamefont {P.~A.}\ \bibnamefont {Shade}}, \
  and\ \bibinfo {author} {\bibfnamefont {D.~M.}\ \bibnamefont {Dimiduk}},\
  }\href {\doibase 10.1146/annurev-matsci-082908-145422} {\bibfield  {journal}
  {\bibinfo  {journal} {Annu. Rev. Mater. Res.}\ }\textbf {\bibinfo {volume}
  {39}},\ \bibinfo {pages} {361} (\bibinfo {year} {2009})}\BibitemShut
  {NoStop}%
\bibitem [{\citenamefont {Dimiduk}\ \emph {et~al.}(2006)\citenamefont
  {Dimiduk}, \citenamefont {Woodward}, \citenamefont {LeSar},\ and\
  \citenamefont {Uchic}}]{dimiduk06}%
  \BibitemOpen
  \bibfield  {author} {\bibinfo {author} {\bibfnamefont {D.~M.}\ \bibnamefont
  {Dimiduk}}, \bibinfo {author} {\bibfnamefont {C.}~\bibnamefont {Woodward}},
  \bibinfo {author} {\bibfnamefont {R.}~\bibnamefont {LeSar}}, \ and\ \bibinfo
  {author} {\bibfnamefont {M.~D.}\ \bibnamefont {Uchic}},\ }\href@noop {}
  {\bibfield  {journal} {\bibinfo  {journal} {Science}\ }\textbf {\bibinfo
  {volume} {312}},\ \bibinfo {pages} {1188} (\bibinfo {year}
  {2006})}\BibitemShut {NoStop}%
\bibitem [{\citenamefont {Csikor}\ \emph {et~al.}(2007)\citenamefont {Csikor},
  \citenamefont {Motz}, \citenamefont {Weygand}, \citenamefont {Zaiser},\ and\
  \citenamefont {Zapperi}}]{csikor07}%
  \BibitemOpen
  \bibfield  {author} {\bibinfo {author} {\bibfnamefont {F.~F.}\ \bibnamefont
  {Csikor}}, \bibinfo {author} {\bibfnamefont {C.}~\bibnamefont {Motz}},
  \bibinfo {author} {\bibfnamefont {D.}~\bibnamefont {Weygand}}, \bibinfo
  {author} {\bibfnamefont {M.}~\bibnamefont {Zaiser}}, \ and\ \bibinfo {author}
  {\bibfnamefont {S.}~\bibnamefont {Zapperi}},\ }\href@noop {} {\bibfield
  {journal} {\bibinfo  {journal} {Science}\ }\textbf {\bibinfo {volume}
  {318}},\ \bibinfo {pages} {251} (\bibinfo {year} {2007})}\BibitemShut
  {NoStop}%
\bibitem [{\citenamefont {Friedman}\ \emph {et~al.}(2012)\citenamefont
  {Friedman}, \citenamefont {Jennings}, \citenamefont {Tsekenis}, \citenamefont
  {Kim}, \citenamefont {Tao}, \citenamefont {Uhl}, \citenamefont {Greer},\ and\
  \citenamefont {Dahmen}}]{Dahmen_2012}%
  \BibitemOpen
  \bibfield  {author} {\bibinfo {author} {\bibfnamefont {N.}~\bibnamefont
  {Friedman}}, \bibinfo {author} {\bibfnamefont {A.~T.}\ \bibnamefont
  {Jennings}}, \bibinfo {author} {\bibfnamefont {G.}~\bibnamefont {Tsekenis}},
  \bibinfo {author} {\bibfnamefont {J.~Y.}\ \bibnamefont {Kim}}, \bibinfo
  {author} {\bibfnamefont {M.}~\bibnamefont {Tao}}, \bibinfo {author}
  {\bibfnamefont {J.~T.}\ \bibnamefont {Uhl}}, \bibinfo {author} {\bibfnamefont
  {J.~R.}\ \bibnamefont {Greer}}, \ and\ \bibinfo {author} {\bibfnamefont
  {K.~A.}\ \bibnamefont {Dahmen}},\ }\href
  {https://link.aps.org/doi/10.1103/PhysRevLett.109.095507} {\bibfield
  {journal} {\bibinfo  {journal} {Phys. Rev. Lett.}\ }\textbf {\bibinfo
  {volume} {109}},\ \bibinfo {pages} {095507} (\bibinfo {year}
  {2012})}\BibitemShut {NoStop}%
\bibitem [{\citenamefont {Papanikolaou}\ \emph {et~al.}(2011)\citenamefont
  {Papanikolaou}, \citenamefont {Bohn}, \citenamefont {Sommer}, \citenamefont
  {Durin}, \citenamefont {Zapperi},\ and\ \citenamefont
  {Sethna}}]{sethna_2011}%
  \BibitemOpen
  \bibfield  {author} {\bibinfo {author} {\bibfnamefont {S.}~\bibnamefont
  {Papanikolaou}}, \bibinfo {author} {\bibfnamefont {F.}~\bibnamefont {Bohn}},
  \bibinfo {author} {\bibfnamefont {R.~L.}\ \bibnamefont {Sommer}}, \bibinfo
  {author} {\bibfnamefont {G.}~\bibnamefont {Durin}}, \bibinfo {author}
  {\bibfnamefont {S.}~\bibnamefont {Zapperi}}, \ and\ \bibinfo {author}
  {\bibfnamefont {J.~P.}\ \bibnamefont {Sethna}},\ }\href {\doibase
  10.1038/nphys1884} {\bibfield  {journal} {\bibinfo  {journal} {Nat. Phys.}\
  }\textbf {\bibinfo {volume} {7}},\ \bibinfo {pages} {316} (\bibinfo {year}
  {2011})}\BibitemShut {NoStop}%
\bibitem [{\citenamefont {Bak}\ \emph {et~al.}(1987)\citenamefont {Bak},
  \citenamefont {Tang},\ and\ \citenamefont {Wiesenfeld}}]{Wiesenfeld_1987}%
  \BibitemOpen
  \bibfield  {author} {\bibinfo {author} {\bibfnamefont {P.}~\bibnamefont
  {Bak}}, \bibinfo {author} {\bibfnamefont {C.}~\bibnamefont {Tang}}, \ and\
  \bibinfo {author} {\bibfnamefont {K.}~\bibnamefont {Wiesenfeld}},\ }\href
  {\doibase 10.1103/PhysRevLett.59.381} {\bibfield  {journal} {\bibinfo
  {journal} {Phys. Rev. Lett.}\ }\textbf {\bibinfo {volume} {59}},\ \bibinfo
  {pages} {381} (\bibinfo {year} {1987})}\BibitemShut {NoStop}%
\bibitem [{\citenamefont {Nix}\ and\ \citenamefont {Lee}(2011)}]{Nix_Lee_2011}%
  \BibitemOpen
  \bibfield  {author} {\bibinfo {author} {\bibfnamefont {W.~D.}\ \bibnamefont
  {Nix}}\ and\ \bibinfo {author} {\bibfnamefont {S.-W.}\ \bibnamefont {Lee}},\
  }\href {\doibase 10.1080/14786435.2010.502141} {\bibfield  {journal}
  {\bibinfo  {journal} {Philos. Mag.}\ }\textbf {\bibinfo {volume} {91}},\
  \bibinfo {pages} {1084} (\bibinfo {year} {2011})}\BibitemShut {NoStop}%
\bibitem [{\citenamefont {Ryu}\ \emph {et~al.}(2013)\citenamefont {Ryu},
  \citenamefont {Nix},\ and\ \citenamefont {Cai}}]{ryu13}%
  \BibitemOpen
  \bibfield  {author} {\bibinfo {author} {\bibfnamefont {I.}~\bibnamefont
  {Ryu}}, \bibinfo {author} {\bibfnamefont {W.~D.}\ \bibnamefont {Nix}}, \ and\
  \bibinfo {author} {\bibfnamefont {W.}~\bibnamefont {Cai}},\ }\href@noop {}
  {\bibfield  {journal} {\bibinfo  {journal} {Acta Mater.}\ }\textbf {\bibinfo
  {volume} {61}},\ \bibinfo {pages} {3233} (\bibinfo {year}
  {2013})}\BibitemShut {NoStop}%
\bibitem [{\citenamefont {Ryu}\ \emph {et~al.}(2015)\citenamefont {Ryu},
  \citenamefont {Cai}, \citenamefont {Nix},\ and\ \citenamefont
  {Gao}}]{nix_gao_2015}%
  \BibitemOpen
  \bibfield  {author} {\bibinfo {author} {\bibfnamefont {I.}~\bibnamefont
  {Ryu}}, \bibinfo {author} {\bibfnamefont {W.}~\bibnamefont {Cai}}, \bibinfo
  {author} {\bibfnamefont {W.~D.}\ \bibnamefont {Nix}}, \ and\ \bibinfo
  {author} {\bibfnamefont {H.}~\bibnamefont {Gao}},\ }\href {\doibase
  10.1016/j.actamat.2015.05.032} {\bibfield  {journal} {\bibinfo  {journal}
  {Acta Mater.}\ }\textbf {\bibinfo {volume} {95}},\ \bibinfo {pages} {176}
  (\bibinfo {year} {2015})}\BibitemShut {NoStop}%
\bibitem [{\citenamefont {Man}\ \emph {et~al.}(2009)\citenamefont {Man},
  \citenamefont {Obrtlik},\ and\ \citenamefont {Pol\'ak}}]{man09}%
  \BibitemOpen
  \bibfield  {author} {\bibinfo {author} {\bibfnamefont {J.}~\bibnamefont
  {Man}}, \bibinfo {author} {\bibfnamefont {K.}~\bibnamefont {Obrtlik}}, \ and\
  \bibinfo {author} {\bibfnamefont {J.}~\bibnamefont {Pol\'ak}},\ }\href
  {\doibase 10.1080/14786430902917616} {\bibfield  {journal} {\bibinfo
  {journal} {Philos. Mag.}\ }\textbf {\bibinfo {volume} {89}},\ \bibinfo
  {pages} {1295} (\bibinfo {year} {2009})}\BibitemShut {NoStop}%
\bibitem [{\citenamefont {Levitin}\ and\ \citenamefont
  {Loskutov}(2009)}]{levitin09}%
  \BibitemOpen
  \bibfield  {author} {\bibinfo {author} {\bibfnamefont {V.}~\bibnamefont
  {Levitin}}\ and\ \bibinfo {author} {\bibfnamefont {S.}~\bibnamefont
  {Loskutov}},\ }\href@noop {} {\emph {\bibinfo {title} {Strained Metallic
  Surfaces}}}\ (\bibinfo  {publisher} {Wiley-VCH},\ \bibinfo {address}
  {Weinheim},\ \bibinfo {year} {2009})\BibitemShut {NoStop}%
\bibitem [{\citenamefont {Goto}\ \emph {et~al.}(2008)\citenamefont {Goto},
  \citenamefont {Han}, \citenamefont {Yakushiji}, \citenamefont {Kim},\ and\
  \citenamefont {Lim}}]{goto08}%
  \BibitemOpen
  \bibfield  {author} {\bibinfo {author} {\bibfnamefont {M.}~\bibnamefont
  {Goto}}, \bibinfo {author} {\bibfnamefont {S.}~\bibnamefont {Han}}, \bibinfo
  {author} {\bibfnamefont {T.}~\bibnamefont {Yakushiji}}, \bibinfo {author}
  {\bibfnamefont {S.}~\bibnamefont {Kim}}, \ and\ \bibinfo {author}
  {\bibfnamefont {C.}~\bibnamefont {Lim}},\ }\href@noop {} {\bibfield
  {journal} {\bibinfo  {journal} {Int. J. Fatigue}\ }\textbf {\bibinfo {volume}
  {30}},\ \bibinfo {pages} {1333} (\bibinfo {year} {2008})}\BibitemShut
  {NoStop}%
\bibitem [{\citenamefont {Laurent}\ \emph {et~al.}(2011)\citenamefont
  {Laurent}, \citenamefont {Tantawi}, \citenamefont {Dolgashev}, \citenamefont
  {Nantista}, \citenamefont {Higashi}, \citenamefont {Aicheler}, \citenamefont
  {Heikkinen},\ and\ \citenamefont {Wuensch}}]{laurent11}%
  \BibitemOpen
  \bibfield  {author} {\bibinfo {author} {\bibfnamefont {L.}~\bibnamefont
  {Laurent}}, \bibinfo {author} {\bibfnamefont {S.}~\bibnamefont {Tantawi}},
  \bibinfo {author} {\bibfnamefont {V.}~\bibnamefont {Dolgashev}}, \bibinfo
  {author} {\bibfnamefont {C.}~\bibnamefont {Nantista}}, \bibinfo {author}
  {\bibfnamefont {Y.}~\bibnamefont {Higashi}}, \bibinfo {author} {\bibfnamefont
  {M.}~\bibnamefont {Aicheler}}, \bibinfo {author} {\bibfnamefont
  {S.}~\bibnamefont {Heikkinen}}, \ and\ \bibinfo {author} {\bibfnamefont
  {W.}~\bibnamefont {Wuensch}},\ }\href {\doibase
  10.1103/PhysRevSTAB.14.041001} {\bibfield  {journal} {\bibinfo  {journal}
  {Phys. Rev. Accel. Beams}\ }\textbf {\bibinfo {volume} {14}},\ \bibinfo
  {pages} {041001} (\bibinfo {year} {2011})}\BibitemShut {NoStop}%
\bibitem [{\citenamefont {Gardiner}(2004)}]{gardiner04}%
  \BibitemOpen
  \bibfield  {author} {\bibinfo {author} {\bibfnamefont {C.~W.}\ \bibnamefont
  {Gardiner}},\ }\href@noop {} {\emph {\bibinfo {title} {Handbook of Stochastic
  Methods}}}\ (\bibinfo  {publisher} {Springer},\ \bibinfo {address} {Berlin},\
  \bibinfo {year} {2004})\BibitemShut {NoStop}%
\bibitem [{sup()}]{supplemental_material}%
  \BibitemOpen
  \href@noop {} {}\bibinfo {note} {See Supplemental Material, which includes
  Refs.
  \cite{weertman64,greenman67,nadgornyi88,hirth82,ashkenazy03,chatterton66,wang04,descoeudres09,taylor34,progress80},
  at for details of the physical model and additional results from the analysis
  and simulation.}\BibitemShut {Stop}%
\bibitem [{Note2()}]{Note2}%
  \BibitemOpen
  \bibinfo {note} {The parameters have been expressed in this form so that we
  can later conveniently apply a WKB-like perturbation theory with respect to
  $n_c \gg 1$ \cite {dykman94}.}\BibitemShut {Stop}%
\bibitem [{\citenamefont {Assaf}\ and\ \citenamefont
  {Meerson}(2017)}]{assaf17}%
  \BibitemOpen
  \bibfield  {author} {\bibinfo {author} {\bibfnamefont {M.}~\bibnamefont
  {Assaf}}\ and\ \bibinfo {author} {\bibfnamefont {B.}~\bibnamefont
  {Meerson}},\ }\href@noop {} {\bibfield  {journal} {\bibinfo  {journal} {J.
  Phys. A}\ }\textbf {\bibinfo {volume} {50}},\ \bibinfo {pages} {263001}
  (\bibinfo {year} {2017})}\BibitemShut {NoStop}%
\bibitem [{Note3()}]{Note3}%
  \BibitemOpen
  \bibinfo {note} {Note that, since $n_* = O(1)$ is an attracting fixed point
  of Eq. (\ref {eq_rho}), starting from any $n = O(1)$ in Eq. (\ref
  {eq_tau_exact}) leads to the same quantitative result.}\BibitemShut {Stop}%
\bibitem [{Note4()}]{Note4}%
  \BibitemOpen
  \bibinfo {note} {Here, once the system has already settled into a long-lived
  metastable state centered about $n_*$, the metastable distribution slowly
  decays at a rate of $1 / \tau $, representing the BD
  probability.}\BibitemShut {Stop}%
\bibitem [{\citenamefont {Dykman}\ \emph {et~al.}(1994)\citenamefont {Dykman},
  \citenamefont {Mori}, \citenamefont {Ross},\ and\ \citenamefont
  {Hunt}}]{dykman94}%
  \BibitemOpen
  \bibfield  {author} {\bibinfo {author} {\bibfnamefont {M.}~\bibnamefont
  {Dykman}}, \bibinfo {author} {\bibfnamefont {E.}~\bibnamefont {Mori}},
  \bibinfo {author} {\bibfnamefont {J.}~\bibnamefont {Ross}}, \ and\ \bibinfo
  {author} {\bibfnamefont {P.}~\bibnamefont {Hunt}},\ }\href@noop {} {\bibfield
   {journal} {\bibinfo  {journal} {J. Chem. Phys.}\ }\textbf {\bibinfo {volume}
  {100}},\ \bibinfo {pages} {5735} (\bibinfo {year} {1994})}\BibitemShut
  {NoStop}%
\bibitem [{\citenamefont {Assaf}\ and\ \citenamefont
  {Meerson}(2006)}]{assaf06}%
  \BibitemOpen
  \bibfield  {author} {\bibinfo {author} {\bibfnamefont {M.}~\bibnamefont
  {Assaf}}\ and\ \bibinfo {author} {\bibfnamefont {B.}~\bibnamefont
  {Meerson}},\ }\href@noop {} {\bibfield  {journal} {\bibinfo  {journal} {Phys.
  Rev. Lett.}\ }\textbf {\bibinfo {volume} {97}},\ \bibinfo {pages} {200602}
  (\bibinfo {year} {2006})}\BibitemShut {NoStop}%
\bibitem [{\citenamefont {Escudero}\ and\ \citenamefont
  {Kamenev}(2009)}]{escudero09}%
  \BibitemOpen
  \bibfield  {author} {\bibinfo {author} {\bibfnamefont {C.}~\bibnamefont
  {Escudero}}\ and\ \bibinfo {author} {\bibfnamefont {A.}~\bibnamefont
  {Kamenev}},\ }\href@noop {} {\bibfield  {journal} {\bibinfo  {journal} {Phys.
  Rev. E}\ }\textbf {\bibinfo {volume} {79}},\ \bibinfo {pages} {041149}
  (\bibinfo {year} {2009})}\BibitemShut {NoStop}%
\bibitem [{\citenamefont {Assaf}\ and\ \citenamefont
  {Meerson}(2010)}]{assaf10}%
  \BibitemOpen
  \bibfield  {author} {\bibinfo {author} {\bibfnamefont {M.}~\bibnamefont
  {Assaf}}\ and\ \bibinfo {author} {\bibfnamefont {B.}~\bibnamefont
  {Meerson}},\ }\href@noop {} {\bibfield  {journal} {\bibinfo  {journal} {Phys.
  Rev. E}\ }\textbf {\bibinfo {volume} {81}},\ \bibinfo {pages} {021116}
  (\bibinfo {year} {2010})}\BibitemShut {NoStop}%
\bibitem [{Note5()}]{Note5}%
  \BibitemOpen
  \bibinfo {note} {Here we assume $\pi _{n > n_c} \simeq 0$ at $t \ll \tau
  $.}\BibitemShut {Stop}%
\bibitem [{Note6()}]{Note6}%
  \BibitemOpen
  \bibinfo {note} {When $E \simeq E_c$, the term $A_1 \sim E^2$ dominates the
  exponent. However, in the experimental regime of interest, where the BD is
  fluctuation driven, all terms in Eq. (\ref {eq_terms_in_exponent}) are of
  similar magnitude.}\BibitemShut {Stop}%
\bibitem [{Note7()}]{Note7}%
  \BibitemOpen
  \bibinfo {note} {While the $E^2$ provides a reasonable fit to experimentally
  available data at room temperature \cite {nordlund12}, a linear dependence
  provides a better approximation over a wide range of electric fields and
  temperatures \cite {supplemental_material}.}\BibitemShut {Stop}%
\bibitem [{\citenamefont {Cahill}\ \emph {et~al.}(2017)\citenamefont {Cahill},
  \citenamefont {Dolgashev}, \citenamefont {Rosenzweig}, \citenamefont
  {Tantawi},\ and\ \citenamefont {Weathersby}}]{cahill17}%
  \BibitemOpen
  \bibfield  {author} {\bibinfo {author} {\bibfnamefont {A.}~\bibnamefont
  {Cahill}}, \bibinfo {author} {\bibfnamefont {V.}~\bibnamefont {Dolgashev}},
  \bibinfo {author} {\bibfnamefont {J.}~\bibnamefont {Rosenzweig}}, \bibinfo
  {author} {\bibfnamefont {S.}~\bibnamefont {Tantawi}}, \ and\ \bibinfo
  {author} {\bibfnamefont {S.}~\bibnamefont {Weathersby}},\ }in\ \href@noop {}
  {\emph {\bibinfo {booktitle} {Proc. of International Particle Accelerator
  Conference (IPAC'17), Copenhagen, Denmark, 14 19 May, 2017}}},\ \bibinfo
  {series and number} {\bibinfo {series} {International Particle Accelerator
  Conference}\ No.~\bibinfo {number} {8}}\ (\bibinfo  {publisher} {JACoW},\
  \bibinfo {address} {Geneva},\ \bibinfo {year} {2017})\ pp.\ \bibinfo {pages}
  {4395--4398}\BibitemShut {NoStop}%
\bibitem [{wue()}]{wuensch17}%
  \BibitemOpen
  \href@noop {} {}\bibinfo {note} {W. Wuensch (private
  communication)}\BibitemShut {NoStop}%
\bibitem [{Note8()}]{Note8}%
  \BibitemOpen
  \bibinfo {note} {$E_a \simeq 0.1$ is the lowest estimate to date of the
  activation energy of mobile dislocation nucleation \cite
  {zhu08}.}\BibitemShut {Stop}%
\bibitem [{\citenamefont {Zhu}\ \emph {et~al.}(2008)\citenamefont {Zhu},
  \citenamefont {Li}, \citenamefont {Samanta}, \citenamefont {Leach},\ and\
  \citenamefont {Gall}}]{zhu08}%
  \BibitemOpen
  \bibfield  {author} {\bibinfo {author} {\bibfnamefont {T.}~\bibnamefont
  {Zhu}}, \bibinfo {author} {\bibfnamefont {J.}~\bibnamefont {Li}}, \bibinfo
  {author} {\bibfnamefont {A.}~\bibnamefont {Samanta}}, \bibinfo {author}
  {\bibfnamefont {A.}~\bibnamefont {Leach}}, \ and\ \bibinfo {author}
  {\bibfnamefont {K.}~\bibnamefont {Gall}},\ }\href {\doibase
  10.1103/PhysRevLett.100.025502} {\bibfield  {journal} {\bibinfo  {journal}
  {Phys. Rev. Lett.}\ }\textbf {\bibinfo {volume} {100}},\ \bibinfo {pages}
  {025502} (\bibinfo {year} {2008})}\BibitemShut {NoStop}%
\bibitem [{\citenamefont {Bonneville}\ \emph {et~al.}(1988)\citenamefont
  {Bonneville}, \citenamefont {Escaig},\ and\ \citenamefont
  {Martin}}]{bonneville88}%
  \BibitemOpen
  \bibfield  {author} {\bibinfo {author} {\bibfnamefont {J.}~\bibnamefont
  {Bonneville}}, \bibinfo {author} {\bibfnamefont {B.}~\bibnamefont {Escaig}},
  \ and\ \bibinfo {author} {\bibfnamefont {J.}~\bibnamefont {Martin}},\
  }\href@noop {} {\bibfield  {journal} {\bibinfo  {journal} {Acta Metall.}\
  }\textbf {\bibinfo {volume} {36}},\ \bibinfo {pages} {1989} (\bibinfo {year}
  {1988})}\BibitemShut {NoStop}%
\bibitem [{\citenamefont {Couteau}\ \emph {et~al.}(2011)\citenamefont
  {Couteau}, \citenamefont {Kruml},\ and\ \citenamefont {Martin}}]{couteau11}%
  \BibitemOpen
  \bibfield  {author} {\bibinfo {author} {\bibfnamefont {O.}~\bibnamefont
  {Couteau}}, \bibinfo {author} {\bibfnamefont {T.}~\bibnamefont {Kruml}}, \
  and\ \bibinfo {author} {\bibfnamefont {J.~L.}\ \bibnamefont {Martin}},\
  }\href@noop {} {\bibfield  {journal} {\bibinfo  {journal} {Acta Mater.}\
  }\textbf {\bibinfo {volume} {59}},\ \bibinfo {pages} {4207} (\bibinfo {year}
  {2011})}\BibitemShut {NoStop}%
\bibitem [{\citenamefont {Ryu}\ \emph {et~al.}(2011)\citenamefont {Ryu},
  \citenamefont {Kang},\ and\ \citenamefont {Cai}}]{ryu11}%
  \BibitemOpen
  \bibfield  {author} {\bibinfo {author} {\bibfnamefont {S.}~\bibnamefont
  {Ryu}}, \bibinfo {author} {\bibfnamefont {K.}~\bibnamefont {Kang}}, \ and\
  \bibinfo {author} {\bibfnamefont {W.}~\bibnamefont {Cai}},\ }\href@noop {}
  {\bibfield  {journal} {\bibinfo  {journal} {Proc. Natl. Acad. Sci. U.S.A.}\
  }\textbf {\bibinfo {volume} {108}},\ \bibinfo {pages} {5174} (\bibinfo {year}
  {2011})}\BibitemShut {NoStop}%
\bibitem [{\citenamefont {Scheffer}\ \emph {et~al.}(2009)\citenamefont
  {Scheffer}, \citenamefont {Bascompte}, \citenamefont {Brock}, \citenamefont
  {Brovkin}, \citenamefont {Carpenter}, \citenamefont {Dakos}, \citenamefont
  {Held}, \citenamefont {van Nes}, \citenamefont {Rietkerk},\ and\
  \citenamefont {Sugihara}}]{scheffer09}%
  \BibitemOpen
  \bibfield  {author} {\bibinfo {author} {\bibfnamefont {M.}~\bibnamefont
  {Scheffer}}, \bibinfo {author} {\bibfnamefont {J.}~\bibnamefont {Bascompte}},
  \bibinfo {author} {\bibfnamefont {W.}~\bibnamefont {Brock}}, \bibinfo
  {author} {\bibfnamefont {V.}~\bibnamefont {Brovkin}}, \bibinfo {author}
  {\bibfnamefont {S.}~\bibnamefont {Carpenter}}, \bibinfo {author}
  {\bibfnamefont {V.}~\bibnamefont {Dakos}}, \bibinfo {author} {\bibfnamefont
  {H.}~\bibnamefont {Held}}, \bibinfo {author} {\bibfnamefont {E.}~\bibnamefont
  {van Nes}}, \bibinfo {author} {\bibfnamefont {M.}~\bibnamefont {Rietkerk}}, \
  and\ \bibinfo {author} {\bibfnamefont {G.}~\bibnamefont {Sugihara}},\
  }\href@noop {} {\bibfield  {journal} {\bibinfo  {journal} {Nature (London)}\
  }\textbf {\bibinfo {volume} {461}},\ \bibinfo {pages} {53} (\bibinfo {year}
  {2009})}\BibitemShut {NoStop}%
\bibitem [{\citenamefont {Weiss}\ \emph {et~al.}(2007)\citenamefont {Weiss},
  \citenamefont {Richeton}, \citenamefont {Louchet}, \citenamefont {Chmelik},
  \citenamefont {Dobron}, \citenamefont {Entemeyer}, \citenamefont {Lebyodkin},
  \citenamefont {Lebedkina}, \citenamefont {Fressengeas},\ and\ \citenamefont
  {McDonald}}]{weiss07}%
  \BibitemOpen
  \bibfield  {author} {\bibinfo {author} {\bibfnamefont {J.}~\bibnamefont
  {Weiss}}, \bibinfo {author} {\bibfnamefont {T.}~\bibnamefont {Richeton}},
  \bibinfo {author} {\bibfnamefont {F.}~\bibnamefont {Louchet}}, \bibinfo
  {author} {\bibfnamefont {F.}~\bibnamefont {Chmelik}}, \bibinfo {author}
  {\bibfnamefont {P.}~\bibnamefont {Dobron}}, \bibinfo {author} {\bibfnamefont
  {D.}~\bibnamefont {Entemeyer}}, \bibinfo {author} {\bibfnamefont
  {M.}~\bibnamefont {Lebyodkin}}, \bibinfo {author} {\bibfnamefont
  {T.}~\bibnamefont {Lebedkina}}, \bibinfo {author} {\bibfnamefont
  {C.}~\bibnamefont {Fressengeas}}, \ and\ \bibinfo {author} {\bibfnamefont
  {R.~J.}\ \bibnamefont {McDonald}},\ }\href {\doibase
  10.1103/PhysRevB.76.224110} {\bibfield  {journal} {\bibinfo  {journal} {Phys.
  Rev. B}\ }\textbf {\bibinfo {volume} {76}},\ \bibinfo {pages} {224110}
  (\bibinfo {year} {2007})}\BibitemShut {NoStop}%
\bibitem [{\citenamefont {Degiovanni}\ \emph {et~al.}(2016)\citenamefont
  {Degiovanni}, \citenamefont {Wuensch},\ and\ \citenamefont
  {Giner~Navarro}}]{degiovanni16}%
  \BibitemOpen
  \bibfield  {author} {\bibinfo {author} {\bibfnamefont {A.}~\bibnamefont
  {Degiovanni}}, \bibinfo {author} {\bibfnamefont {W.}~\bibnamefont {Wuensch}},
  \ and\ \bibinfo {author} {\bibfnamefont {J.}~\bibnamefont {Giner~Navarro}},\
  }\href {\doibase 10.1103/PhysRevAccelBeams.19.032001} {\bibfield  {journal}
  {\bibinfo  {journal} {Phys. Rev. Accel. Beams}\ }\textbf {\bibinfo {volume}
  {19}},\ \bibinfo {pages} {032001} (\bibinfo {year} {2016})}\BibitemShut
  {NoStop}%
\bibitem [{\citenamefont {Wuensch}(2013)}]{wuensch13}%
  \BibitemOpen
  \bibfield  {author} {\bibinfo {author} {\bibfnamefont {W.}~\bibnamefont
  {Wuensch}},\ }\href@noop {} {\emph {\bibinfo {title} {Advances in the
  understanding of the physical processes of vacuum breakdown}}},\ \bibinfo
  {type} {Tech. Rep.}\ \bibinfo {number} {CERN-OPEN-2014-028}\ (\bibinfo
  {institution} {CERN},\ \bibinfo {address} {Geneva},\ \bibinfo {year}
  {2013})\BibitemShut {NoStop}%
\bibitem [{\citenamefont {Weertman}\ and\ \citenamefont
  {Weertman}(1964)}]{weertman64}%
  \BibitemOpen
  \bibfield  {author} {\bibinfo {author} {\bibfnamefont {J.}~\bibnamefont
  {Weertman}}\ and\ \bibinfo {author} {\bibfnamefont {J.~R.}\ \bibnamefont
  {Weertman}},\ }\href@noop {} {\emph {\bibinfo {title} {Elementary Dislocation
  Theory}}}\ (\bibinfo  {publisher} {MacMillan},\ \bibinfo {address} {New
  York},\ \bibinfo {year} {1964})\BibitemShut {NoStop}%
\bibitem [{\citenamefont {Greenman}\ \emph {et~al.}(1967)\citenamefont
  {Greenman}, \citenamefont {Vreeland},\ and\ \citenamefont
  {Wood}}]{greenman67}%
  \BibitemOpen
  \bibfield  {author} {\bibinfo {author} {\bibfnamefont {W.~F.}\ \bibnamefont
  {Greenman}}, \bibinfo {author} {\bibfnamefont {T.}~\bibnamefont {Vreeland},
  \bibfnamefont {Jr.}}, \ and\ \bibinfo {author} {\bibfnamefont {D.~S.}\
  \bibnamefont {Wood}},\ }\href {\doibase 10.1063/1.1710178} {\bibfield
  {journal} {\bibinfo  {journal} {J. Appl. Phys.}\ }\textbf {\bibinfo {volume}
  {38}},\ \bibinfo {pages} {3595} (\bibinfo {year} {1967})}\BibitemShut
  {NoStop}%
\bibitem [{\citenamefont {Nadgornyi}(1988)}]{nadgornyi88}%
  \BibitemOpen
  \bibfield  {author} {\bibinfo {author} {\bibfnamefont {E.}~\bibnamefont
  {Nadgornyi}},\ }\href {\doibase https://doi.org/10.1016/0079-6425(88)90005-9}
  {\bibfield  {journal} {\bibinfo  {journal} {Prog. Mater. Sci.}\ }\textbf
  {\bibinfo {volume} {31}},\ \bibinfo {pages} {1 } (\bibinfo {year}
  {1988})}\BibitemShut {NoStop}%
\bibitem [{\citenamefont {Hirth}\ and\ \citenamefont {Lothe}(1982)}]{hirth82}%
  \BibitemOpen
  \bibfield  {author} {\bibinfo {author} {\bibfnamefont {J.~P.}\ \bibnamefont
  {Hirth}}\ and\ \bibinfo {author} {\bibfnamefont {J.}~\bibnamefont {Lothe}},\
  }\href@noop {} {\emph {\bibinfo {title} {Theory of Dislocations}}}\ (\bibinfo
   {publisher} {Krieger},\ \bibinfo {address} {Malabar, FL},\ \bibinfo {year}
  {1982})\BibitemShut {NoStop}%
\bibitem [{\citenamefont {Mordehai}\ \emph {et~al.}(2003)\citenamefont
  {Mordehai}, \citenamefont {Ashkenazy}, \citenamefont {Kelson},\ and\
  \citenamefont {Makov}}]{ashkenazy03}%
  \BibitemOpen
  \bibfield  {author} {\bibinfo {author} {\bibfnamefont {D.}~\bibnamefont
  {Mordehai}}, \bibinfo {author} {\bibfnamefont {Y.}~\bibnamefont {Ashkenazy}},
  \bibinfo {author} {\bibfnamefont {I.}~\bibnamefont {Kelson}}, \ and\ \bibinfo
  {author} {\bibfnamefont {G.}~\bibnamefont {Makov}},\ }\href {\doibase
  10.1103/PhysRevB.67.024112} {\bibfield  {journal} {\bibinfo  {journal} {Phys.
  Rev. B}\ }\textbf {\bibinfo {volume} {67}},\ \bibinfo {pages} {024112}
  (\bibinfo {year} {2003})}\BibitemShut {NoStop}%
\bibitem [{\citenamefont {Chatterton}(1966)}]{chatterton66}%
  \BibitemOpen
  \bibfield  {author} {\bibinfo {author} {\bibfnamefont {P.}~\bibnamefont
  {Chatterton}},\ }\href@noop {} {\bibfield  {journal} {\bibinfo  {journal}
  {Proc. Phys. Soc.}\ }\textbf {\bibinfo {volume} {88}},\ \bibinfo {pages}
  {231} (\bibinfo {year} {1966})}\BibitemShut {NoStop}%
\bibitem [{\citenamefont {Wang}\ \emph {et~al.}(2004)\citenamefont {Wang},
  \citenamefont {Wang}, \citenamefont {He}, \citenamefont {Xu},\ and\
  \citenamefont {Li}}]{wang04}%
  \BibitemOpen
  \bibfield  {author} {\bibinfo {author} {\bibfnamefont {X.}~\bibnamefont
  {Wang}}, \bibinfo {author} {\bibfnamefont {M.}~\bibnamefont {Wang}}, \bibinfo
  {author} {\bibfnamefont {P.}~\bibnamefont {He}}, \bibinfo {author}
  {\bibfnamefont {Y.}~\bibnamefont {Xu}}, \ and\ \bibinfo {author}
  {\bibfnamefont {Z.}~\bibnamefont {Li}},\ }\href@noop {} {\bibfield  {journal}
  {\bibinfo  {journal} {J. Appl. Phys.}\ }\textbf {\bibinfo {volume} {96}},\
  \bibinfo {pages} {6752} (\bibinfo {year} {2004})}\BibitemShut {NoStop}%
\bibitem [{\citenamefont {Descoeudres}\ \emph {et~al.}(2009)\citenamefont
  {Descoeudres}, \citenamefont {Levinsen}, \citenamefont {Calatroni},
  \citenamefont {Taborelli},\ and\ \citenamefont {Wuensch}}]{descoeudres09}%
  \BibitemOpen
  \bibfield  {author} {\bibinfo {author} {\bibfnamefont {A.}~\bibnamefont
  {Descoeudres}}, \bibinfo {author} {\bibfnamefont {Y.}~\bibnamefont
  {Levinsen}}, \bibinfo {author} {\bibfnamefont {S.}~\bibnamefont {Calatroni}},
  \bibinfo {author} {\bibfnamefont {M.}~\bibnamefont {Taborelli}}, \ and\
  \bibinfo {author} {\bibfnamefont {W.}~\bibnamefont {Wuensch}},\ }\href@noop
  {} {\bibfield  {journal} {\bibinfo  {journal} {Phys. Rev. Accel. Beams}\
  }\textbf {\bibinfo {volume} {12}},\ \bibinfo {pages} {092001} (\bibinfo
  {year} {2009})}\BibitemShut {NoStop}%
\bibitem [{\citenamefont {Taylor}(1934)}]{taylor34}%
  \BibitemOpen
  \bibfield  {author} {\bibinfo {author} {\bibfnamefont {G.~I.}\ \bibnamefont
  {Taylor}},\ }\href@noop {} {\bibfield  {journal} {\bibinfo  {journal} {Proc.
  R. Soc. A}\ }\textbf {\bibinfo {volume} {145}},\ \bibinfo {pages} {362}
  (\bibinfo {year} {1934})}\BibitemShut {NoStop}%
\bibitem [{\citenamefont {Sevillano}\ \emph {et~al.}(1980)\citenamefont
  {Sevillano}, \citenamefont {van Houtte},\ and\ \citenamefont
  {Aernoudt}}]{progress80}%
  \BibitemOpen
  \bibfield  {author} {\bibinfo {author} {\bibfnamefont {J.}~\bibnamefont
  {Sevillano}}, \bibinfo {author} {\bibfnamefont {P.}~\bibnamefont {van
  Houtte}}, \ and\ \bibinfo {author} {\bibfnamefont {E.}~\bibnamefont
  {Aernoudt}},\ }\href {\doibase
  http://dx.doi.org/10.1016/0079-6425(80)90002-X} {\bibfield  {journal}
  {\bibinfo  {journal} {Prog. Mater. Sci.}\ }\textbf {\bibinfo {volume} {25}},\
  \bibinfo {pages} {135 } (\bibinfo {year} {1980})}\BibitemShut {NoStop}%
\end{thebibliography}%

\appendix

\section{\LARGE Supplemental Material}

\subsection{Creation and Depletion Rates}

In this section we describe the considerations leading to the
formulation of the kinetic equations,
leading up to Eq. (1) in the main text.

The rate at which new mobile dislocations
are created in a slip plane is determined by the density of sources,
which is proportional to the density of in-plane pre-existing defects $c$
and the mobile dislocation density $\rho$.
Both of these quantities represent in-plane densities,
and therefore are measured in units of $1 / \text{nm}$.
Based on the amount of defects we see in experimental samples,
we assign $c$ a value of 1 $\mu$m\textsuperscript{-1}.

In addition,
the rate is proportional to the rate of creation of dislocations by each defect.
Since the creation of dislocations is thermally activated,
the rate should be inversely proportional to a temperature-dependent factor
$\exp[(E_a - \Omega\sigma) / (k_B T)]$,
divided by the creation time of each dislocation.
Here $E_a$ and $\Omega$ are the activation energy and volume, respectively,
of a dislocation nucleation source,
whose values we estimate in the main text,
while $\sigma$ is the surface stress of the metal.

To find the attempt frequency of the nucleation, we consider, for convenience,
a Frank-Read type source.
The creation time of a dislocation in such a source is $t = L/v$,
with $L$ the length of the source, and $v$ the velocity of the dislocation.
The threshold stress needed to activate a source of length $L$ or longer is
$\sigma_{th} = 2Gb/L$,
where $G = 48$ GPa is the shear modulus,
and $b = 0.25$ nm is the Burgers vector \cite{weertman64}.
If the amount of sources decreases rapidly as a function of length,
then $L \approx 2Gb/\sigma$.
For stresses ranging from 0.2 \cite{greenman67, nadgornyi88}
up to 400 MPa \cite{hirth82, ashkenazy03},
the dislocation velocity in Cu is approximately a linear function of $\sigma$,
$v = 50 C_t \sigma/G$,
where $C_t = 2.31\times 10^3$ m/s
is the propagation velocity of sound in Cu \cite{hirth82}.
Therefore $t = G^2b / (25 C_t \sigma^2)$, giving us a total creation rate
\begin{equation}
  \dot{\rho}^+ =
  \frac{25\kappa C_t}{G^2 b} (\rho + c) \sigma^2 e^{-\frac{E_a - \Omega\sigma}{k_B T}},
\end{equation}
where $\kappa$ is a kinetic factor,
combining the fraction of
mobile dislocations that have been pinned
and therefore contribute nucleation sources,
and the activation entropy of the sources \cite{ryu11}.
We estimate the value of $\kappa$ in the main text.

Mobile dislocations can be depleted by interactions
with other mobile dislocations
as well as with pre-existing defects,
and ejection to the surface.
Assuming that the last mechanism is considerably slower than the first two,
we can write the depletion rate as $\dot{\rho}^- = \xi\rho v (c + \rho)$.
Here $\xi$ is a dimensionless proportionality factor,
representing trap efficiency.
For simplicity, we assign it a value of 1.
Substituting once again for $v$ we have
\begin{equation}
  \dot{\rho}^- = \frac{50\xi C_t}{G} \sigma\rho(c + \rho).
\end{equation}

The stress in metal subjected to an external electric field is
composed of the Maxwell stress
due to the external field $E$ applied,
and of the average internal stress created by the dislocations themselves.  
The Maxwell stress is $\epsilon_0 (\beta E)^2 / 2$,
with the dimensionless parameter $\beta$ representing
the ratio of the
effective electric field at the mobile dislocation nucleation site to $E$.
$\beta$ is expected to depend both on surface geometry
and the electric field distribution.
Specifically, $\beta$ is expected to vary with $\rho$,
since it depends on the aspect ratio of protrusions created on the surface
\cite{chatterton66, wang04, descoeudres09}.
However, it can be shown by numerical analysis that in the regime of interest,
as determined by the values of the rest of the parameters,
the aspect ratio remains nearly constant until breakdown.
Therefore, we consider it to be a constant over time,
whose value we estimate in the main text, for every given cathode geometry.

The second term of the stress is proportional to $Gb/d$,
where $d$ is the average distance between dislocations
\cite{taylor34, progress80}.
In the experimental setups examined in the main text,
a pulsed electric field is applied,
and the breakdown rate (BDR) is constant over time.
Since there is no memory effect, 
we assume a constant sessile dislocation population
whose contribution to the total stress from all slip planes saturates.
As a result, we take into consideration
only the stress caused by the mobile dislocations,
whose density varies over time.
In multi-slip plane systems $d$ is proportional to $\rho^{-1/2}$,
with $\rho$ measured in units of nm\textsuperscript{-2}
\cite{taylor34, progress80}.
However, when considering only one slip plane as in our model,
we expect the relation to be $d \sim \rho^{-1}$,
with $\rho$ in units of nm\textsuperscript{-1}, as described above
(and also see below).
We therefore find that, altogether, the stress is
$\sigma = \epsilon_0 (\beta E)^2 / 2 + ZGb\rho$,
where the dimensionless parameter $Z$, in the second term of the stress,
is a structural parameter linking the stress to the dislocation density.
For simplicity, we assign it a value of 1.

\begin{figure}
  \centering
  \includegraphics[width=8.5cm]{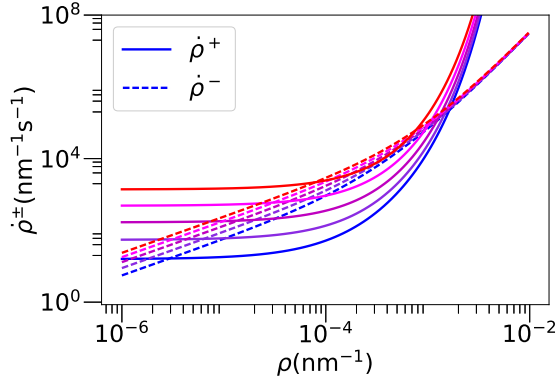}
  \caption
  {
    \label{fig_analytical_drhomp}
    $\dot{\rho}^+$ and $\dot{\rho}^-$ for five electric fields (bottom to top):
    150, 190, 230, 270, and 310 MV/m.
  }
\end{figure}

Defining new constants of the form $\alpha \equiv \Omega / (k_B T)$,
$A_1 \equiv \epsilon_0 (\beta E)^2 / 2$, $a_2 \equiv ZGb$,
$B_1 \equiv 25\kappa C_t e^{-\frac{E_a}{k_B T}} / (G^2 b)$,
and $b_2 \equiv 50\xi C_t / G$,
we arrive at Eq. (1) in the main text.
The values of $\beta$ = 4.8$\pm$0.1, $\kappa$ = 0.41$\pm$0.02,
$\Omega$ = 5.4$\pm$0.2 eV/GPa, and $E_a$ = 0.08$\pm$0.002 eV,
found by the fitting procedure in the main text,
give us the following values for the constants:
$A_1$ = 100 Pa\,(MV/m)\textsuperscript{-2}$E^2$, $a_2$ = 12 GPa\,nm,
$B_1$ = 1850 MPa\textsuperscript{-2}\,s\textsuperscript{-1},
$b_2$ = 2410 m\,GPa\textsuperscript{-1}\,s\textsuperscript{-1},
$c$ = 1 $\mu$m\textsuperscript{-1}, $\alpha$ = 210 GPa\textsuperscript{-1}.
Figure \ref{fig_analytical_drhomp} shows the values of $\dot{\rho}^+$
and $\dot{\rho}^-$ for the nominal values of the above parameters.

\subsection{Linear Approximation of the BDR}

\begin{figure}
  \centering
  \includegraphics[width=8.5cm]{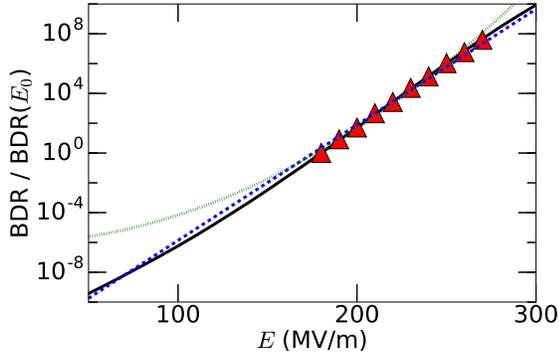}
  \caption
  {
    \label{fig_linear_vs_square}
    BDR as a function of the electric field.
    The solid line is the metastable approximation
    [Eq. (9) in the main text],
    the triangles are the simulation results,
    and the dashed and dotted lines are linear and quadratic fits,
    respectively.
  }
\end{figure}

For practical purposes it is useful to identify a simple function
of the dependence of the BDR on the electric field,
which can be later fitted to experimental results.
Our results strongly indicate that a linear fit of the logarithm serves
as a good approximation over a wide range of electric fields,
see Eq. (11) in the main text and Fig. \ref{fig_linear_vs_square}
(for fields between 50 and 300 MV/m).
Note that the previously suggested dependence,
$\tau \sim \exp(\alpha E^2)$ \cite{nordlund12},
agrees with our model within the range of currently available data
but significantly diverges from our results outside that range.
However, we stress that our model includes a discernibly different
temperature-dependent term,
which can be distinguished by dedicated experiments.

\subsection{Variance of the QSD}

\begin{figure}
  \centering
  \includegraphics[width=8.5cm]{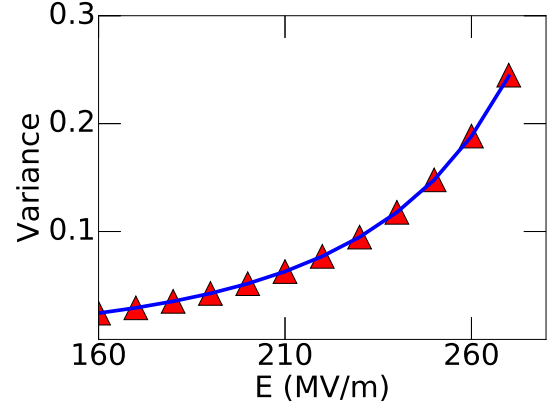}
  \caption
  {
    \label{fig_variance}
    Variance of the QSD as function of electric field:
    analytical result (lines), extracted from Eq. (7) in the main text,
    and simulation (triangles).
  }
\end{figure}

Figure \ref{fig_variance} shows the variance of the
quasi-stationary distribution (QSD) as a function of the electric field,
for the same set of parameters as in Fig. \ref{fig_analytical_drhomp}.
Here the simulation results agree well with a numerical calculation
of the variance of the theoretical QSD [Eq. (7) in the main text].
As expected, for stronger fields the variance is larger,
and the probability for breakdown increases,
since the system moves more rapidly towards higher values of $n$.
This increase in variance may be experimentally detected
through increased variation in related signals
such as the acoustic emission signal from moving dislocations
within the electrodes,
as well as in the dark current produced between them under increasing field.

\subsection{Volume Density Model}

The model developed in this manuscript discusses in-plane
mobile dislocation density fluctuations,
neglecting interactions between slip planes.
The mobile dislocation density $\rho$ is therefore
a two-dimensional density, measured in units of length per area,
nm\textsuperscript{-1}.
If we were to define $\rho$ as the volume density of mobile dislocations
in units of length per volume, nm\textsuperscript{-2},
the average distance between dislocations would be propotional to $\rho^{-1/2}$
\cite{taylor34, progress80}.
Then, the stress would be $\sigma = \epsilon_0 (\beta E)^2 / 2 + ZGb\rho^{1/2}$.
The creation and depletion rates would be
\begin{align}
  \dot{\rho}^+ &= \frac{25\kappa C_t}{G^2 b} (\rho + c) \sigma^2
  e^{-\frac{E_a - \Omega\sigma}{k_B T}} \\
  \dot{\rho}^- &= \frac{50\xi C_t}{G} \sigma b\rho(c + \rho) \nonumber
\end{align}
where $c$ = 1 $\mu$m\textsuperscript{-2} is now
the volume density of the constant defects,
while all other constants retain their original meaning.
The factor of $b$ in the depletion term was added in order to correctly describe
the probability of two dislocations interacting,
now in a volume instead of a plane,
assuming that the width of a dislocation is equal to the Burgers vector $b$.

\begin{figure}
  \centering
  \includegraphics[width=8.5cm]{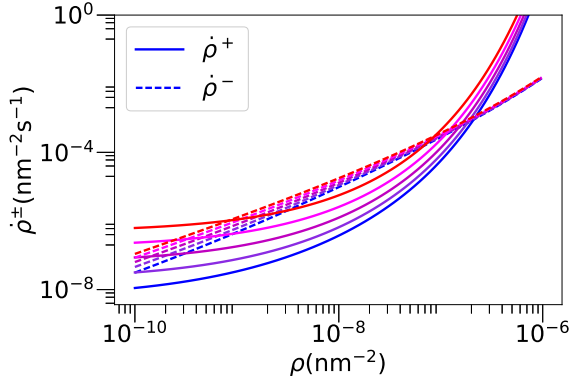}
  \caption
  {
    \label{fig_analytical_drhomp_root_stress}
    $\dot{\rho}^+$ and $\dot{\rho}^-$
    in a model describing volume mobile dislocation density fluctuations,
    for five electric fields (bottom to top):
    150, 190, 230, 270, and 310 MV/m.
  }
\end{figure}

As seen in Fig. \ref{fig_analytical_drhomp_root_stress},
for adjusted values of the parameter set
$\beta$, $\kappa$, $\Omega$, and $E_a$,
$\dot{\rho}^+$ and $\dot{\rho}^-$ as volume density creation and depletion rates
exhibit the same qualitative behavior as in the two-dimensional density model.
The same considerations as in the latter model can then be applied,
once again yielding the $\ln\tau \sim E$ dependence
and BDRs as described in the main text.

\end{document}